\documentclass[aps,showpacs,prl,twocolumn,footinbib]{revtex4-1}
\usepackage[normalem]{ulem}
\usepackage{epsfig}
\usepackage{amsfonts}
\usepackage{amsmath}
\usepackage[utf8]{inputenc}
\usepackage[colorlinks,bookmarks=false,citecolor=blue,linkcolor=blue,urlcolor=blue]{hyperref}

\newcommand{\beq}{\begin{equation}}
\newcommand{\eeq}{\end{equation}}
\newcommand{\bea}{\begin{align}}
\newcommand{\eea}{\end{align}}
\newcommand{\nn}{\nonumber}

\newcommand{\ua}{\uparrow}
\newcommand{\da}{\downarrow}

\begin{document}
\title{Odd-Frequency Triplet Superconductivity at the Helical Edge of a Toplogical Insulator}
\author{Fran\c{c}ois Cr\'epin}
\affiliation{Institute for Theoretical Physics and Astrophysics,
University of W\"urzburg, 97074 W\"urzburg, Germany}
\author{Pablo Burset}
\affiliation{Institute for Theoretical Physics and Astrophysics,
University of W\"urzburg, 97074 W\"urzburg, Germany}
\author{Bj\"orn Trauzettel}
\affiliation{Institute for Theoretical Physics and Astrophysics,
University of W\"urzburg, 97074 W\"urzburg, Germany}

\date{\today}

\begin{abstract}
%



%

Non-local pairing processes at the edge of a two-dimensional topological insulator in proximity to an $s$-wave superconductor are usually suppressed by helicity. However, additional proximity of a ferromagnetic insulator can substantially influence the helical constraint and therefore open a new conduction channel by allowing for crossed Andreev reflection (CAR) processes. We show a one-to-one correspondence between CAR and the emergence of odd-frequency triplet superconductivity. Hence, non-local transport experiments that identify CAR in helical liquids yield smoking-gun evidence for unconventional superconductivity. Interestingly, we identify a setup -- composed of a superconductor flanked by two ferromagnetic insulators -- that allows us to favor CAR over electron cotunneling which is known to be a difficult but essential task to be able to measure CAR.


%
\end{abstract}


\maketitle

{\it Introduction.}--- The influence of strong spin-orbit coupling and the constraint of time reversal symmetry are responsible for the appearance of helical electronic channels at the edge of two-dimensional (2D) topological insulators~\cite{Kane05, Kane05b,Zhang06b}. Ample evidence for the experimental detection of these edge states has been seen in transport \cite{Konig07,Knez11,Brune12} and scanning SQUID experiments \cite{Nowack13}. Proximity of an ordinary $s$-wave superconductor can induce pairing at the helical edge~\cite{Fu08, Fu09b}. Interestingly, the interplay of helicity and superconducting order gives rise to unconventional proximity-induced superconductivity (SC) in these systems. However, it is fair to say that it is very difficult to unambiguously probe the emergence of unconventional SC because -- often times -- conventional and unconventional signatures of SC look alike. Several recent experiments have demonstrated (as a first step towards the detection of unconventional SC) that helical liquids as boundary states of quantum spin Hall systems can indeed be brought in proximity to $s$-wave superconductors~\cite{Knez12} and serve, for instance, as conducting channels of a Josephson junction~\cite{Hart14,Pribiag14}.

Importantly, helicity guarantees perfect local Andreev reflection \cite{Adroguer10} -- the conversion of an electron into a hole with opposite spin -- at the interface between the normal and the proximity-induced superconducting region called NS junction. Evidently, perfect local Andreev reflection can give rise to sub-gap Andreev states, for instance, in Josephson junction setups. Among these sub-gap states, the one with zero excitation energy is its own charge-conjugate state. It has been coined Majorana (bound) state and has attracted a lot of attention because of potential applications of this anyonic state for topological quantum computing \cite{Nayak08}. If a ferromagnet (FM) is placed in vicinity to the NS interface the Majorana (bound) state can be localized in the region between the FM and the SC~\cite{Fu09b}.  This localization allows us to make a formal connection to the physics of a finite-size 1D $p$-wave topological superconductor~\cite{Kitaev01}, and its potential realizations in spin-orbit nanowires~\cite{Lutchyn10, Oreg10b}.



\begin{figure}[h]
\centering
\includegraphics[width=0.47\textwidth,clip]{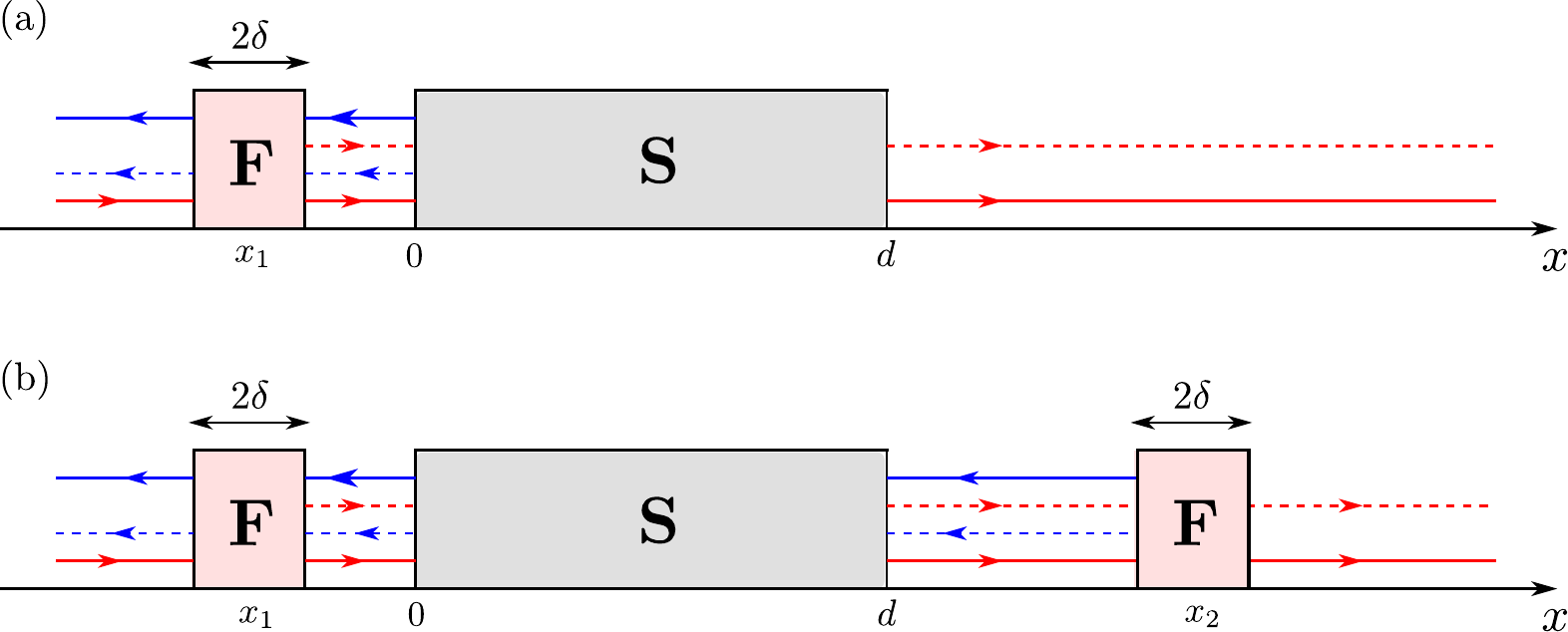}
\caption{(Color online) Examples of hybrid junctions in a helical liquid. A possible scattering process is represented, with a spin up right moving electron (solid red line), a spin down left moving electron (solid blue line), a spin down right moving hole (red dashed line) and a spin up left moving hole (blue dashed line).}
\label{Fig:system}
\end{figure}

In this Letter, we give for the first time a complete characterization of the pairing symmetries in a helical liquid, in proximity to both FM insulators and SC. The typical setups we have in mind are sketched in Fig.~\ref{Fig:system}. Interestingly, we discover that the presence of both FM and SC yields an unconventional contribution to the pairing amplitude that is odd (O) in frequency, triplet (T) in spin space, and even (E) in orbital space (in short: OTE). Now, the natural question arises whether it is possible to unambiguously detect this exotic pairing in our setup. Remarkably, the answer is yes. {Indeed, the FM also allows for equal-spin crossed Andreev reflections, that would otherwise be totally suppressed by helicity. Conversely, in the absence of the FM, equal spin OTE pairing is forbidden by time-reversal symmetry and CAR contributions completely vanish.} {Our proposal, based on well established measurements of the differential conductance~\cite{Lohneysen04, Mopurgo05}, turns out to provide the simplest constituents necessary to directly probe odd-frequency triplet pairing. Hence, CAR is a crystal-clear transport signature of unconventional superconductivity in helical liquids. {Our system substantially differs from the original literature about non-helical, diffusive, FSF junctions, where long-range proximity effect was considered as evidence for odd-frequency triplet superconductivity~\cite{ Efetov05, *Birge10, *Eschrig11}.}

Previously, the shot noise related to CAR processes has been proposed as a way to detect Majorana bound states~\cite{Nilsson08,Liu13}. Moreover, in a nanowire setup, local Andreev reflection measurements have been suggested to probe spin-triplet superconducting correlations of Majorana fermions~\cite{Liu15}. {However these works both address} a fundamentally different scientific question than ours. In our Letter, we describe the novel idea of relating CAR processes directly to symmetry properties of the proximity-induced pairing amplitude. This idea is heavily based on the peculiar transport properties of helical edge states.


{\it Model and analysis of pairing symmetries.}--- Our starting point is the Hamiltonian (we set $\hbar =1$)
\beq
H = \frac{1}{2} \int_{-\infty}^\infty dx \; \Psi^\dagger(x) \mathcal{H}_{\textrm{BdG}}(x) \Psi(x)\;,  \label{eq:H_1}
\eeq
where we have introduced the spinor $\Psi(x) = [\psi_\ua(x), \psi_\da(x), \psi_\da^\dagger(x),-\psi_\ua^\dagger(x)]^{\rm{T}}$ and the Bogoliubov-de Gennes (BdG) Hamiltonian $\mathcal{H}_{\textrm{BdG}}(x) = (- i v_F \partial_x \sigma_z -\mu) \tau_z + \mathbf{m}(x)\cdot \boldsymbol{\sigma} +  \Delta(x)\tau_x $. The Pauli matrices $\boldsymbol{\sigma} = (\sigma_x,\sigma_y,\sigma_z)$ and $\boldsymbol{\tau} = (\tau_x,\tau_y,\tau_z)$ act in spin and particle-hole space, respectively. The phase of the pairing potential plays no role in our calculation and we have omitted it for the sake of clarity. In practice we take $\Delta(x) = \theta(x)\theta(d-x) \Delta_0$ and  $\mathbf{m}(x) = \mathbf{m_1}(x)  +  \mathbf{m_2}(x) $ with $\mathbf{m_1}(x) = \theta(x_1+\delta - x)\theta(x-x_1+\delta) (m_1 \cos \lambda_1, m_1 \sin \lambda_1, m_{z,1})$ and  $\mathbf{m_2}(x) = \theta(x - x_2+\delta)\theta(x_2 + \delta - x) (m_2 \cos \lambda_2, m_2 \sin \lambda_2, m_{z,2})$. In Fig.~\ref{Fig:system}(a), we illustrate the simpler case where $m_2 = m_{z,2} = 0$, whereas Fig.~\ref{Fig:system}(b) corresponds to the most general case we consider here. We are interested in the pairing amplitude, that is, the anomalous (electron-hole) part of the retarded Green's function
\beq
\check{G}^{R}(r,r') = - i \theta(t-t') \left\langle \left\{ \Psi(r), \Psi^\dagger(r') \right\} \right\rangle\;, \label{Eq:GR}
\eeq
where $r=(x,t)$ and $r'=(x',t')$. Its Fourier transform $\check{G}^R(x,x',\omega) = \int dt\; e^{i(\omega+i0^+)(t-t')}\check{G}^R(r,r')$ satisfies the equation of motion $\left[ \omega - H_{\rm{BdG}}(x) \right] \check{G}^R(x,x',\omega) = \!\delta(x-x')$ and  is discontinuous in $x=x'$, such that
\beq
\lim_{\varepsilon \to 0} \left[  \check{G}^R(x'+\varepsilon , x',\omega) - \check{G}^R(x'-\varepsilon, x',\omega) \right]= \frac{\sigma_z\tau_z}{iv_F}. \label{Eq:BC_1}
\eeq
For pedagogical reasons, we first investigate the setup of Fig.~\ref{Fig:system}(a) in the limit $d \to +\infty$. We construct the retarded Green's function from scattering states satisfying {\it outgoing wave} boundary conditions~\cite{Kashiwaya00, Herrera10}. These states define an eigenbasis of the BdG Hamiltonian.  For a given energy $\omega$, there are four independent scattering states $\phi_{\omega,1}$ (electron incoming from the left), $\phi_{\omega,2}$ (hole incoming from the left), $\phi_{\omega,3}$ (electron-like quasi-particle incoming from the right), $\phi_{\omega,4}$ (hole-like quasi-particle incoming from the right). Each incoming particle can either be Andreev reflected, with probability amplitude $a_i$, or normal reflected, with probability amplitude $b_i$. We give the asymptotic forms of the scattering states in the supplementary material~\footnote{See the supplementary material for detailed expressions of the
	scattering states and of the components of the Green's function,
	as well as elements of the scattering matrix in the case of a ferromagnetic
	impurity.}. We proceed now to analyze the symmetry content of the pairing in hybrid junctions based on helical liquids. The Green's function is a $4\times 4$ matrix of the general form
\beq
\check{G}^R(x,x',\omega) = \begin{pmatrix}
\hat{G}^R_{ee}(x,x',\omega) &\hat{G}^R_{eh}(x,x',\omega) \\
\hat{G}^R_{he}(x,x',\omega) & \hat{G}^R_{hh}(x,x',\omega) \\
\end{pmatrix}\;.\label{Eq:GR_2}
\eeq
The pairing amplitude, given by the off-diagonal elements of Eq.~\eqref{Eq:GR_2} must be completely antisymmetric under the exchange of positions, times and spin components of the electrons making up the Cooper pair.  The spin symmetry can be made explicit by writing
\beq
\hat{G}_{eh}^{R}(x,x',\omega)  =  f^{R}_0(x,x',\omega) \sigma_0 + f^{R}_i(x,x',\omega) \sigma_i   \;,
\eeq
where $f^{R}_0$ is the singlet (S) component of the pairing amplitude while $f^{R}_{1,2,3}$ make for the three components of the triplet (T). The antisymmetry of the pairing amplitude translates as
\begin{align}
f_0^R(x,x',\omega) &= f_0^A(x',x,-\omega)\;, \\
f_{1,2,3}^R(x,x',\omega) &= - f_{1,2,3}^A(x',x,-\omega)\;,
\end{align}
where $f_{0,1,2,3}^A$ is built from the advanced Green's function $\check{G}^A(x,x',\omega) = \check{G}^R(x',x,\omega)^\dagger$. The rule of thumb is generally that the orbital part of the singlet (resp. triplet) is even (resp. odd). However, this is not the most general case, as the behavior under the exchange of times also contribute to the antisymmetry of the Cooper pair Green's function~\cite{Berezinskii74}. In principle, each of the $f^{R/A}_{j=0,1,2,3}$ can be even (E) or odd (O) under the exchange of $x$ and $x'$, that is, $f^{R}_{j}(x,x',\omega) = \pm f^{R}_{j}(x',x,\omega) $, or under $\omega \rightarrow - \omega$, that is, $f^{R}_{j}(x,x',\omega) = \pm f^{A}_{j}(x,x',-\omega) $. The pairing is thus classified according to these three symmetries and, adopting the notation [frequency, spin, orbital], there are only 4 allowed combinations that comply with the total antisymmetry of the pair amplitude: (ESE, OSO, ETO, OTE)~\cite{Tanaka07, *Tanaka12}. Actually, in the case at hand of an NS junction, translational invariance is broken and the $f^{R/A}_{j=0,1,2,3}$ are necessarily a mixture of functions with even and odd orbital symmetries. We find that the components of the pairing amplitude are of the form~\footnote{In our setups, the interface region is defined as the region in space
	between the FM and the SC, whereas the bulk is the region in space
	below the SC.}
\begin{align}
f^{R/A}_{0}(x,x',\omega) &= f^{R/A}_{0,\textrm{interface}}(x,x',\omega) + f^{R/A}_{0,\textrm{bulk}}(x,x',\omega), \nn \\
f^{R/A}_{3}(x,x',\omega) &= f^{R/A}_{3,\textrm{interface}}(x,x',\omega) + f^{R/A}_{3,\textrm{bulk}}(x,x',\omega),\nn\\
f^{R/A}_{\pm}(x,x',\omega) &= f^{R/A}_{\pm,\textrm{interface}}(x,x',\omega),
\end{align}
where $f_{+} = f_1 - i f_2$ (resp. $f_{-} = f_1 + i f_2$) is simply the $\ua\ua$ (resp. $\da\da$) pairing amplitude. The bulk contributions are all proportional to $e^{-\kappa(\omega)\lvert x -x' \rvert}$ with $\kappa(\omega) = \sqrt{\Delta^2-\omega^2}/v_F$, while the interface contributions decay to zero as $e^{-\kappa(\omega)(x +x')}$~\cite{Note1}. Although the pairing term in the Hamiltonian \eqref{eq:H_1} is of the singlet form and reflects the symmetry of the nearby superconductor~\cite{[{The latter statement has been verified with a self-consistent calculation of the induced gap in }] [{}]Schaffer11b}, a triplet component naturally emerges in the bulk, because of helicity. Deep inside the proximity induced superconducting region, we find a coexistence of a singlet and a triplet  component. We have verified that they belong to the conventional symmetry classes, namely ESE and ETO, respectively~\cite{Yokoyama10,  *Black13}. At the interface however, all of the symmetry classes matter. The singlet has both ESE and OSO components while the triplet $f_3$ has both ETO and OTE components~\cite{Black12}. However, the two other components of the triplet, $f_+$ and $f_-$, corresponding to $\ua\ua$ and $\da\da$ pairing respectively, exist only in the exotic OTE class. Explicitly, we find
\begin{align}
f_{+,\textrm{interface}}^R(x,x',\omega) &=  - \Delta_0 \frac{ e^{(i\mu-\kappa(\omega))(x+x')} }{2v_F\sqrt{\Delta_0^2-\omega^2}}    b_3(\omega) \;, \label{Eq:f+} \\
f_{-,\textrm{interface}}^R(x,x',\omega) &=  - \Delta_0 \frac{e^{(i\mu-\kappa(\omega))(x+x')} }{2v_F\sqrt{\Delta_0^2-\omega^2}} b_4(\omega) \;, \\
f_{+,\textrm{interface}}^A(x,x',\omega) &=  - \Delta_0 \frac{ e^{(i\mu-\kappa(\omega))(x+x')} }{2v_F\sqrt{\Delta_0^2-\omega^2}}    b_4(\omega)^* \;,\\
f_{-,\textrm{interface}}^A(x,x',\omega) &=  - \Delta_0 \frac{e^{(i\mu-\kappa(\omega))(x+x')} }{2v_F\sqrt{\Delta_0^2-\omega^2}} b_3(\omega)^* \;,
\end{align}
with $b_{3/4}(\omega)$ the normal reflection amplitudes for quasi-particles incoming from the right~\cite{Note1}. It is clear that $f^{R/A}_{\pm}(x,x',\omega) =  f^{R/A}_{\pm}(x',x,\omega) $. Furthermore, the fundamental particle-hole symmetry of the BdG Hamiltonian, $U_C \mathcal{H}_{\textrm{BdG}}^*U_C^\dagger= - \mathcal{H}_{\textrm{BdG}}$, with $U_C = \sigma_y\tau_y$, yields  $b_3(\omega) = -b_4(-\omega)^*$. It hence follows that $f^{R}_{\pm}(x,x',\omega) = - f^{A}_{\pm}(x,x',-\omega)$.  In the absence of the ferromagnetic insulator, normal reflection processes are forbidden by helicity and $b_{3/4} \to 0$. Therefore, proximity of a ferromagnet induces odd-frequency triplet pairing in the $\ua\ua$ and $\da \da$ channels,  at the edge of a 2D topological insulator. This constitutes one of the main findings of this Letter. The symmetry classification is summarized in Table~\ref{T:Class}. \\

\begin{table}
\centering
  \begin{tabular}{ | c  | c || c | c | }
    \hline
    & Pairing & Interface & Bulk \\ \hline \hline
    $f_0$ & $\ua\da-\da\ua$ & ESE+OSO & ESE \\ \hline
    $f_3$ & $\ua\da +\da\ua$ & ETO+OTE & ETO \\ \hline
    $f_\pm$ & $\ua\ua$, $\da\da$ &OTE  & X  \\ \hline
    \end{tabular}
        \caption{Symmetry classification of the pairing in a NFS junction at the edge of 2D topological insulator. E and O indicate even and odd, while S and T stand for singlet and triplet. The first letter gives the symmetry with respect to frequency while the last gives the orbital symmetry of the pairing.} \label{T:Class}
\end{table}

\begin{figure}%
    \centering
  \includegraphics[width=0.45\textwidth]{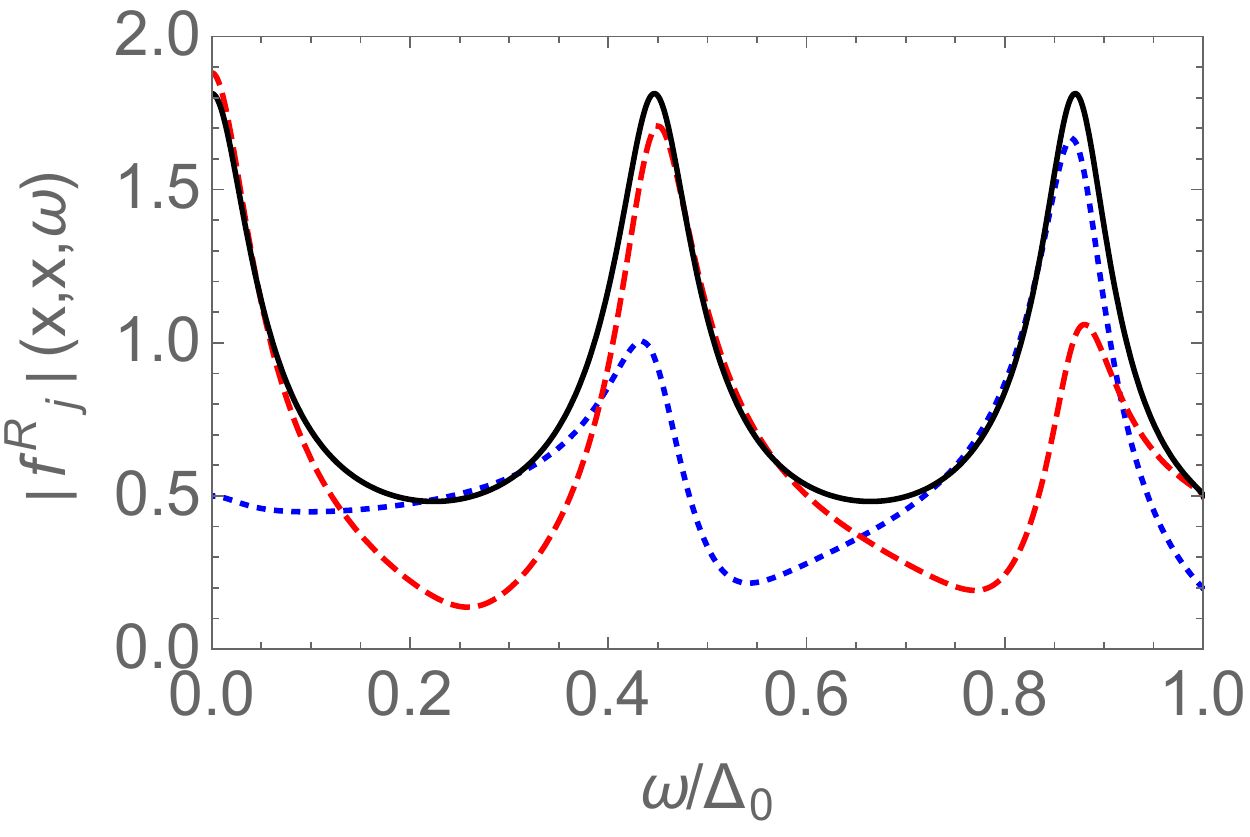} %
    \caption{(Color online) Pairing amplitudes $f_0(x,x,\omega)$ (blue dotted line), $f_3(x,x,\omega)$ (red dashed line) and $f_+(x,x,\omega)$ (black solid line), at the NS interface $x=0$ [see Fig.~\ref{Fig:system}(a)]. The impurity is located at $ x_1= -3\xi$ and its strength is $m_0 = 0.5$. The positions of the peaks correspond to the localized Andreev states in the junction and signal an enhancement of triplet pairing.}%
    \label{Fig:pairing}%
\end{figure}

For simplicity, we now consider the case of a ferromagnetic impurity, meaning that we reduce the width of the ferromagnetic domain $\delta \to 0$ while keeping {the dimensionless quantity} $(m_1 \delta)/v_F = m_0$ finite. We take $m_{z,1} = 0$, with no consequence on the physics~\footnote{In the $z$ direction, the effect of the ferromagnet is just to shift locally the chemical potentials of right and left movers, without inducing spin-flip.}. In that case, we expect the formation of Andreev bound states between the impurity and the superconducting region, the number of which is essentially given by the ratio of the distance to the superconductor $\lvert x_1 \rvert$ and the coherence length $\xi = v_F/\Delta_0$. Their main characteristic is to allow for perfect local Andreev reflection~\cite{Crepin14c}. Interestingly, it is possible to relate the physics of these localized states to the induced triplet superconductivity. In Fig.~\ref{Fig:pairing}, we plot the moduli of the local pairing amplitudes $f_0(x,x,\omega)$, $f_3(x,x,\omega)$, and $f_+(x,x,\omega)$ as a function of the excitation energy $\omega$, for $x=0$ (see Fig.~\ref{Fig:system}(a)). Note that $f_+(x,x',\omega) = f_-(x,x',\omega)$ in the impurity limit. Enforcing $x = x'$ effectively gets rid of the odd orbital components. Therefore we probe only ESE and OTE components of the pairing. We have taken $\lvert x_1 \rvert = 3\xi$ and a quite large impurity strength $m_0 = 0.5$, in order to have three well-defined bound states below the gap. The $\ua\ua$ pairing amplitude is especially strong close to the energies of the bound states. It is actually maximal at the NS interface $x=0$, and decays exponentially into the bulk as can be seen in Eq.~\eqref{Eq:f+}. This confirms the proximity induced character of the odd-frequency triplet pairing in a helical NFS junction. It should be noted however that no local observable can directly probe the  $\ua\ua$ or $\da\da$ pairing. We have indeed checked that the $\ua\ua$ and $\da\da$ local pairing amplitudes vanish at the NF interface and, by continuity, in all of the normal lead: $f^R_\pm(x,x,\omega) = 0$ for $x \leq x_1$. Furthermore, we obtain $f^R_0(x,x,\omega) = f^R_3(x,x,\omega)$ for $x \leq x_1$, that is equal $\ua\da-\da\ua$ and $\ua\da+\da\ua$ correlations in the normal lead, again a direct consequence of helicity. Thus, the local conductance $G_{11}(V_1) = \partial{I_1}/\partial{V_1}$, with $I_1$ the current injected in the left (normal) lead and $V_1$ the bias between this normal lead and the superconductor, does neither probe $\ua\ua$ nor $\da\da$ correlations. In order to detect these correlations, one needs to consider the non-local conductance.



\begin{figure}%
    \centering
  \includegraphics[width=0.47\textwidth]{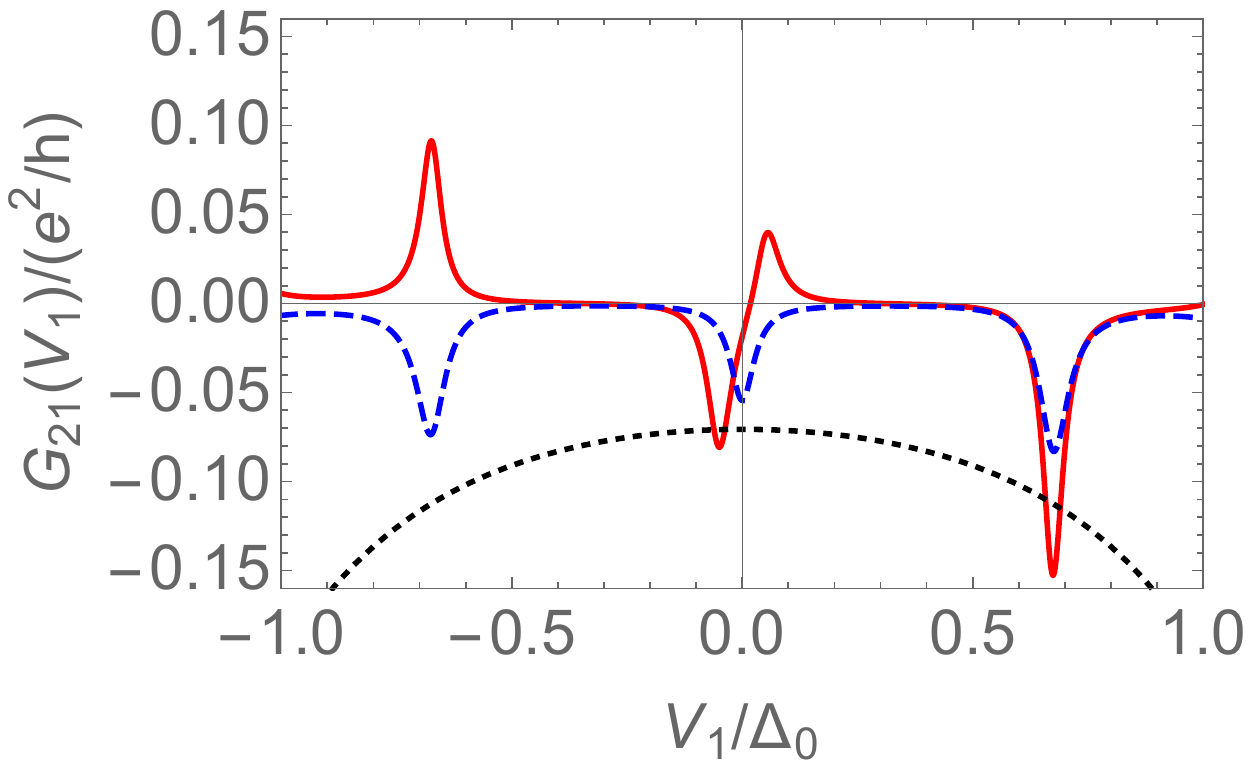} %
    \caption{(Color online) Non-local conductance $G_{21} = -\partial I_2 /\partial V_1 = T_{CAR} - T_{EC}$, at zero temperature, in the setup of Figs.~\ref{Fig:system}(a) (blue dashed line) and (b) (red solid line). A positive signal indicates an excess of crossed Andreev reflection over electron tunneling. We have taken $d = 2\xi$, $\delta = 0.5 \xi$, $m_1 = 1.5\Delta_0$, $m_{z,1} = 0$, $x_1 = -2\xi$, $x_2= d+\xi$, and a non-zero chemical potential $\mu = 0.5 \Delta_0$. In the first setup (blue dashed line) there is no ferromagnetic impurity in the right lead, $m_2 = m_{z,2} = 0$, while in the second setup (red solid line) $m_2 = m_1$ and $m_{z,2} = m_{z,1} = 0$. The black dotted line shows the conductance in the absence of any ferromagnetic impurity. }%
    \label{Fig:CAR}%
\end{figure}

{\it Crossed Andreev reflection and transport signatures.}--- We now consider the setup of Fig.~\ref{Fig:system}(a) with $d$ finite. We have seen that helicity forbids to probe the $\ua \ua$ and $\da \da$ pair correlations locally. Indeed, a right moving electron incoming from the left lead cannot be reflected as left moving hole with the same spin. However, the right moving electron can be converted into a right moving hole with the same spin, in the right lead. The latter process, called crossed Andreev reflection (CAR), corresponds to the injection of a Cooper pair in the $\ua \ua$ channel~\cite{Colin06, Guo14}, and probes directly the exotic pairing induced by the proximity of the FM insulator. { We do not expect that charging effects are important in our setup. The reason is that for a size of the SC region of order $\xi$, the corresponding single particle level spacing should be much larger than the corresponding charging energy. Thus,} we have solved the problem based on scattering theory and computed the CAR and electron cotunneling (EC) probabilities. As can be inferred from Eq.~\eqref{Eq:f+}, both EC and CAR probabilities are suppressed exponentially with the length $d$ of the superconducting region. However CAR is enhanced for excitation energies close to that of the Andreev bound-states, consistent with the enhancement of triplet pairing shown in Fig.~\ref{Fig:pairing}. Unfortunately,  EC is always bigger than CAR in that case which is a rather generic situation to arise. The non-local conductance $G_{21} = -\partial I_2 /\partial V_1 = T_{CAR} - T_{EC}$ directly probes the difference between CAR and EC. It can be seen in Fig.~\ref{Fig:CAR} (blue dashed line) to be always negative for this setup. Contrary to other proposals, we cannot filter spin in order to favor CAR~\cite{Falci01}. Another option is to break the symmetry between particle and hole excitations~\cite{Cayssol08}, yet we find that the chemical potential has virtually no effect on the transmission probabilities in the case of a single impurity.

However, there is a very convenient way out of this dilemma: Adding a second FM to the right lead (setup of Fig.~\ref{Fig:system}(b)) allows to significantly enhance CAR over EC. Andreev bound-states will now also arise in the right lead, their number being given by $(x_2-d)/\xi$.  If the length of the superconductor is not too big compared to $\xi$, the two sets of bound states (left and right of the SC) hybridize, and become very sensitive to the chemical potential. We found that CAR and electron tunneling can then be made quite different in systems with a pronounced left-right asymmetry. As an example, we have taken $x_1 = -2\xi $ and $x_2-d = \xi$. The zero bias peak in the conductance splits under the effect of the hybridization, as can be seen in Fig.~\ref{Fig:CAR}. However, the second bound-state, that is localized in the left lead, has no counterpart to hybridize with. This results in a single peak in the non-local conductance, whose sign can be tuned at will with the bias. A positive signal of $G_{21} = -\partial I_2 /\partial V_1 = T_{CAR} - T_{EC}$ is an unambiguous signature of odd-frequency triplet pairing at the edge of 2D topological superconductors.

{\it Conclusions.}--- Helicity at the edge of a quantum spin Hall system implies perfect spin-to-charge conversion: all right movers have spin up and all left movers spin down. In proximity to an $s$-wave superconductor, helical edge states exhibit unconventional pairing amplitudes. We have carefully characterized the symmetry of the superconducting pairing at the helical edge in the additional presence of ferromagnetic insulators. Interestingly, we have discovered a unique relation between the existence of odd-frequency triplet superconductivity in these systems and crossed Andreev reflection processes due to helicity. Remarkably, we have even identified a possibility to favor crossed Andreev reflection over electron cotunneling by a proper tuning of resonance conditions.

{\it Acknowledgments.}--- We thank T.M. Klapwijk and Y. Tanaka for interesting discussions and acknowledge financial support by the DFG-JST research unit ``Topotronics'', the DFG priority
program SPP1666, the Helmholtz Foundation (VITI), and the ENB graduate school on ``Topological Insulators''.

\bibliography{Top_ins_wurzburg}

\onecolumngrid
\appendix

\section{Supplementary material}

In this supplementary material we give detailed expressions for the (retarded) Green's function. We start by discussing the structure of the Green's function in particle-hole and spin space. We then give explicit formulas for the junction of Fig.~\ref{Fig:system}(a) in the limit $d \to \infty$, and give additional information about the pairing symmetries, as well as analytic expressions for the scattering matrix in the case of a ferromagnetic impurity.

\subsection{A. General structure of the Green's function}
It should be noted that because of the particle-hole symmetry of the BdG Hamiltonian, not all sixteen elements of the Green's function are independent. Indeed we have,
\beq
U_C \mathcal{H}_{\textrm{BdG}}^*U_C^\dagger= - \mathcal{H}_{\textrm{BdG}}\;,
\eeq
and
\beq
U_C \check{G}^R(x,x',\omega)^* U_C^\dagger= - \check{G}^R(x,x',-\omega)\;,
\eeq
with $U_C = \sigma_y\tau_y$. Defining,
\beq
\check{G}^R(x,x',\omega) = \begin{pmatrix}
\hat{G}^R_{ee}(x,x',\omega) & \hat{G}^R_{eh}(x,x',\omega) \\
\hat{G}^R_{he}(x,x',\omega) & \hat{G}^R_{hh}(x,x',\omega) \\
\end{pmatrix}\;,\label{Eq:GR_2b}
\eeq
we find that
\begin{align}
\hat{G}^R_{ee}(x,x',-\omega) &= - \sigma_y \hat{G}^R_{hh}(x,x',\omega)^* \sigma_y\;,\\
\hat{G}^R_{eh}(x,x',-\omega) &=  \sigma_y \hat{G}^R_{he}(x,x',\omega)^* \sigma_y\;.
\end{align}
Therefore, the knowledge of e.g. $\hat{G}^R_{ee}(x,x',\omega)$ and $\hat{G}^R_{eh}(x,x',\omega)$ is enough to characterize the full Green's function, and all the associated physical processes. We construct the Green's function from the scattering states with outgoing boundary conditions. For the sake of completeness we give their asymptotic forms, in the leftmost normal region ($x<x_1-\delta$) and in the superconducting region ($x>0$). For a given energy $\omega$, there are four independent scattering states $\phi_{1,\omega}, \phi_{2,\omega}, \phi_{3,\omega}, \phi_{4,\omega}$. In the normal region,
\begin{align}
\phi_{1,\omega}(x) &=  \phi_e^{(+)} e^{i k_e x} + a_1 \phi_h^{(-)} e^{i k_h x} + b_1 \phi_e^{(-)} e^{- i k_e x} \;, \\
\phi_{2,\omega}(x) &=  \phi_h^{(+)} e^{-i k_h x} + b_2 \phi_h^{(-)} e^{i k_h x} + a_2 \phi_e^{(-)} e^{- i k_e x} \;, \\
\phi_{3,\omega}(x) &=   c_3 \phi_e^{(-)} e^{-i k_e x} + d_3 \phi_h^{(-)} e^{ i k_h x} \;, \\
\phi_{4,\omega}(x) &=   d_4 \phi_e^{(-)} e^{-i k_e x} + c_4 \phi_h^{(-)} e^{ i k_h x} \;,
\end{align}
where we have introduced $k_{e/h} = (\mu \pm \omega)/v_F$, as well as $\phi_e^{(+)} = (1,0,0,0)^T$, $\phi_e^{(-)} = (0,1,0,0)^T$, $\phi_h^{(+)} = (0,0,0,1)^T$, and $\phi_h^{(-)} = (0,0,1,0)^T$. In the superconducting region we have
 \begin{align}
\phi_{1,\omega}(x) &=  c_1 \chi_e^{(+)} e^{i k^S_e x} + d_1 \chi_h^{(+)} e^{- i k^S_h x} \;, \\
\phi_{2,\omega}(x) &=  d_2 \chi_e^{(+)} e^{i k^S_e x} + c_2 \chi_h^{(+)} e^{- i k^S_h x} \;, \\
\phi_{3,\omega}(x) &=  \chi_e^{(-)} e^{-i k^S_e x} + a_3 \chi_h^{(+)} e^{-i k^S_h x} + b_3 \chi_e^{(+)} e^{ i k^S_e x}  \label{Eq:scatt1}\\
\phi_{4,\omega}(x) &=   \chi_h^{(-)} e^{-i k^S_h x} + b_4 \chi_h^{(+)} e^{-i k^S_h x} + a_4 \chi_e^{(+)} e^{ i k^S_e x} \label{Eq:scatt2}
\end{align}
where now $k^S_{e/h} = (\mu \pm \sqrt{\omega^2-\Delta^2})/v_F$, for excitation energies above the gap. We also have introduced $\chi_e^{(+)} = \frac{1}{\sqrt{N_e(\omega)}}(e^{A(\omega)},0,1,0)^T$, $\chi_e^{(-)} = \frac{1}{\sqrt{N_e(\omega)}}(0,  e^{A(\omega)},0,1)^T$,  $\chi_h^{(+)} = \frac{1}{\sqrt{N_h(\omega)}}(0,  e^{-A(\omega)},0,1)^T$, $\chi_h^{(-)} = \frac{1}{\sqrt{N_h(\omega)}}(e^{-A(\omega)},0,1,0)^T$, with $N_{e/h}(\omega) = 1 + e^{\pm2A(\omega)}$ as well as $A(\omega) = \textrm{arcosh}(\omega/\Delta_0)$. At a later stage, these expressions will be analytically continued to energies below the gap. The coefficients $(a_i, b_i, c_i, d_i)_{i=1,2,3,4}$  in the scattering states depend on the excitation energy $\omega$ and are the elements of the scattering matrix.  They are obtained by matching the wave-functions at the boundaries between regions. We demand that $\check{G}^R(x,x',\omega)$ satisfies {\it outgoing wave} boundary conditions and construct it in the following way:
\beq
\check{G}^R(x,x',\omega)=
\left\{
\begin{array}{ll}
\! \phi_3(x) A_3(x')^{\rm{T}} + \phi_4(x) A_4(x')^{\rm{T}}    & \!\! \mbox{ if $x<x'$}\\
\! \phi_1(x) A_1(x')^{\rm{T}} + \phi_2(x) A_2(x')^{\rm{T}}    & \!\! \mbox{ if $x>x'$} \nn
\end{array}
\right. \label{Eq:def_GR}
\eeq
The 16 unknowns in the vectors $A_1, A_2, A_3, A_4$ are solutions of Eq.~\eqref{Eq:BC_1}, and are related, in the end, to the elements of the scattering matrix. We find $\hat{G}^R_{ee}(x,x',\omega)$ and $\hat{G}^R_{eh}(x,x',\omega)$ in the various regions of the setup pictured in Fig.~\ref{Fig:system}(a). Given our choice of the spinor  $\Psi(x) = [\psi_\ua(x), \psi_\da(x), \psi_\da^\dagger(x),-\psi_\ua^\dagger(x)]^{\rm{T}}$, the spin components of electron and holes are distributed in the following way:
\beq
\hat{G}^R_{ee} = \begin{pmatrix}
[G^R_{ee}]_{\ua\ua} & [G^R_{ee}]_{\ua\da} \\
[G^R_{ee}]_{\da\ua} & [G^R_{ee}]_{\da\da} \\
\end{pmatrix}\;.
\eeq
for the regular part, and
\beq
\hat{G}^R_{eh} = \begin{pmatrix}
[G^R_{eh}]_{\ua\da} & [G^R_{eh}]_{\ua\ua} \\
[G^R_{eh}]_{\da\da} & [G^R_{eh}]_{\da\ua} \\
\end{pmatrix}\;.
\eeq
for the anomalous part.

\subsection{B. Green's function in the left lead}
 In the normal lead ($x < x_1-\delta$),
\begin{align}
[G^R_{ee}]_{\ua\ua}(\omega,x,x') &=  e^{i k_e (x-x')}\theta(x-x')\;,\\
[G^R_{ee}]_{\da\da}(\omega,x,x') &=   e^{-i k_e (x-x')}\theta(x'-x)\;,\\
[G^R_{ee}]_{\ua\da}(\omega,x,x') &= 0 \;,\\
[G^R_{ee}]_{\da\ua}(\omega,x,x') &=  e^{-i k_e (x+x')} b_1(\omega) \;,
\end{align}
while
\begin{align}
[G^R_{eh}]_{\ua\da}(\omega,x,x') &= 0 \;,\\
[G^R_{eh}]_{\da\ua}(\omega,x,x') &=  e^{i(k_h x'-k_e x)} a_2(\omega)\;,\\
[G^R_{eh}]_{\ua\ua}(\omega,x,x') &=  0\;,\\
[G^R_{eh}]_{\da\da}(\omega,x,x') &=   0\;,
\end{align}
where $k_{e/h} = (\mu \pm \omega)/v_F$, $b_1(\omega)$ is the normal reflection amplitude for electrons and $a_2(\omega)$ is the Andreev reflection amplitude for holes. The Heaviside functions in the diagonal elements of $\hat{G}^R_{ee}(x,x',\omega)$ arise from the discontinuity of the Green's function at $x = x'$ given in Eq.~\eqref{Eq:BC_1}, a direct consequence of dealing with Dirac fermions. Note that some matrix elements in $\hat{G}^R_{ee}(x,x',\omega)$ and $\hat{G}^R_{eh}(x,x',\omega)$ vanish because of helicity, and can never be non-zero, even in the presence of the ferromagnetic impurity. At this level we can already discuss the spin symmetry of the pairing amplitude by writing
\beq
G_{eh}^{R}(x,x',\omega)  =  f^{R}_0(x,x',\omega) \sigma_0 + f^{R}_i(x,x',\omega) \sigma_i   \;. \label{Eq:supp_decomp}
\eeq
It follows directly from the structure of  $\hat{G}_{eh}^{R}(x,x',\omega)$  that $f_1(x,x',\omega) = f_2(x,x',\omega) = 0$ and  $f_0(x,x',\omega) = - f_3(x,x',\omega) $ for all values of $x$ and $x'$ in the left lead. From this we deduce that $\ua \ua$ and $\da\da$ cannot be probed locally, in a transport experiment, and that local Andreev reflection in a helical liquid induces equal singlet ($\ua \da - \da \ua$)  and triplet ($\ua \da + \da\ua$)  correlations.

\subsection{C. Green's function in the superconductor}
In the region covered by the superconductor ($x>0$) the physics is much richer, and we find, for excitation energies below the gap
\begin{align}
[G^R_{ee}]_{\ua\ua}(\omega,x,x') &=  \frac{1}{i v_F} \frac{e^{i \mu (x-x')}}{e^{i\eta(\omega)}-e^{-i\eta(\omega)}} \left[ e^{-\kappa(\omega)\lvert x -x' \rvert}\left(\theta(x-x')e^{i\eta(\omega)}+\theta(x'-x)e^{-i\eta(\omega)}\right) + a_4(\omega)  e^{-\kappa(\omega)(x+x')}\right]\;, \\
[G^R_{ee}]_{\da\da}(\omega,x,x') &= \frac{1}{i v_F} \frac{e^{-i \mu (x-x')}}{e^{i\eta(\omega)}-e^{-i\eta(\omega)}} \left[ e^{-\kappa(\omega)\lvert x -x' \rvert}\left(\theta(x-x')e^{-i\eta(\omega)}+\theta(x'-x)e^{i\eta(\omega)}\right) + a_3(\omega)  e^{-\kappa(\omega)(x+x')}\right]\;,\\
[G^R_{ee}]_{\ua\da}(\omega,x,x') &=  \frac{1}{i v_F} \frac{e^{i \mu (x+x')}}{e^{i\eta(\omega)}-e^{-i\eta(\omega)}}  b_3(\omega) e^{-\kappa(\omega)(x+x')}\;,\\
[G^R_{ee}]_{\da\ua}(\omega,x,x') &=  \frac{1}{i v_F} \frac{e^{-i \mu (x+x')}}{e^{i\eta(\omega)}-e^{-i\eta(\omega)}}  b_4(\omega) e^{-\kappa(\omega)(x+x')}\;,
\end{align}
for the regular part, and
\begin{align}
[G^R_{eh}]_{\ua\da}(\omega,x,x') &=  \frac{1}{i v_F} \frac{e^{i \mu (x-x')}}{e^{i\eta(\omega)}-e^{-i\eta(\omega)}} \left[ e^{-\kappa(\omega)\lvert x -x' \rvert} + a_4(\omega) e^{i\eta(\omega)} e^{-\kappa(\omega)(x+x')}\right]\;, \label{Eq:GR_1}\\
[G^R_{eh}]_{\da\ua}(\omega,x,x') &=  \frac{1}{i v_F} \frac{e^{-i \mu (x-x')}}{e^{i\eta(\omega)}-e^{-i\eta(\omega)}} \left[ e^{-\kappa(\omega)\lvert x -x' \rvert} + a_3(\omega) e^{-i\eta(\omega)} e^{-\kappa(\omega)(x+x')}\right]\;,\\
[G^R_{eh}]_{\ua\ua}(\omega,x,x') &=  \frac{1}{i v_F} \frac{e^{i \mu (x+x')}}{e^{i\eta(\omega)}-e^{-i\eta(\omega)}}  b_3(\omega) e^{-\kappa(\omega)(x+x')}\;,\\
[G^R_{eh}]_{\da\da}(\omega,x,x') &=  \frac{1}{i v_F} \frac{e^{-i \mu (x+x')}}{e^{i\eta(\omega)}-e^{-i\eta(\omega)}}  b_4(\omega) e^{-\kappa(\omega)(x+x')}\;,
\end{align}
for the anomalous part, with $\kappa(\omega)= \sqrt{\Delta_0^2-\omega^2}$ and $\eta(\omega) = \arccos(\omega/\Delta_0)$. Each of the matrix elements is the sum of a {\it bulk} term (proportional to $e^{-\kappa(\omega)\lvert x -x' \rvert}$) and an {\it interface} term (proportional to $e^{-\kappa(\omega)(x+x')}$).  In order to discuss the symmetries of the pairing amplitude, we also give the anomalous part of the advanced Green's function
\begin{align}
[G^A_{eh}]_{\ua\da}(\omega,x,x') &=  \frac{1}{i v_F} \frac{e^{i \mu (x-x')}}{e^{i\eta(\omega)}-e^{-i\eta(\omega)}} \left[ e^{-\kappa(\omega)\lvert x -x' \rvert} + a_4(\omega)^* e^{i\eta(\omega)} e^{-\kappa(\omega)(x+x')}\right]\;,\\
[G^A_{eh}]_{\da\ua}(\omega,x,x') &=  \frac{1}{i v_F} \frac{e^{-i \mu (x-x')}}{e^{i\eta(\omega)}-e^{-i\eta(\omega)}} \left[ e^{-\kappa(\omega)\lvert x -x' \rvert} + a_3(\omega)^* e^{-i\eta(\omega)} e^{-\kappa(\omega)(x+x')}\right]\;,\\
[G^A_{eh}]_{\ua\ua}(\omega,x,x') &=  \frac{1}{i v_F} \frac{e^{i \mu (x+x')}}{e^{i\eta(\omega)}-e^{-i\eta(\omega)}}  b_4(\omega)^* e^{-\kappa(\omega)(x+x')}\;,\\
[G^A_{eh}]_{\da\da}(\omega,x,x') &=  \frac{1}{i v_F} \frac{e^{-i \mu (x+x')}}{e^{i\eta(\omega)}-e^{-i\eta(\omega)}}  b_3(\omega)^* e^{-\kappa(\omega)(x+x')}\;.
\end{align}
We expect the pairing amplitude (the Cooper pair wave-function) to be totally antisymmetric under exchange of positions, spins and times of the two electrons making up the pair. More precisely
\begin{align}
[G^R_{eh}]_{\ua\ua}(\omega,x,x')  &= - [G^A_{eh}]_{\ua\ua}(-\omega,x',x)\;, \label{Eq:supp_upup}\\
 [G^R_{eh}]_{\da\da}(\omega,x,x')  &= - [G^A_{eh}]_{\da\da}(-\omega,x',x)\;, \\
 [G^R_{eh}]_{\ua\da}(\omega,x,x')  &=  [G^A_{eh}]_{\da\ua}(-\omega,x',x)\;.  \label{Eq:supp_updown}
\end{align}
Note that the last equality comes without a minus sign because of the choice of basis we have made. The spinor defining the Hamiltonian of Eq.~\eqref{eq:H_1} indeed reads $\Psi(x) = [\psi_\ua(x), \psi_\da(x), \psi_\da^\dagger(x),-\psi_\ua^\dagger(x)]^{\rm{T}}$. The minus sign appearing in the definition of the spinor implicitly takes care of the minus sign that should arise upon exchanging $\ua$ and $\da$ in Eq.~\eqref{Eq:supp_updown}. From particle-hole symmetry, we have $a_4(\omega) = - a_3(-\omega)^*$, $b_4(\omega) = - b_3(-\omega)^*$, and with the help of $e^{i\eta(-\omega)} = -e^{-i\eta(\omega)}$ one can check the general validity of Eqs.~\eqref{Eq:supp_upup} to \eqref{Eq:supp_updown}. In order to isolate the spin symmetry we use the decomposition of Eq.~\eqref{Eq:supp_decomp}. We have, on the retarded side,
\begin{align}
f_{0,\textrm{bulk}}^R(x,x',\omega) &=  -  \frac{ \Delta_0 }{2v_F\sqrt{\Delta_0^2-\omega^2}}  e^{-\kappa(\omega)\lvert x -x' \rvert} \cos[\mu(x-x')]\;, \label{Eq:0_bulk} \\
f_{0,\textrm{interface}}^R(x,x',\omega) &=  -  \frac{\Delta_0 }{4v_F\sqrt{\Delta_0^2-\omega^2}}  e^{-\kappa(\omega)(x+x')}\left[  e^{i \mu (x-x')} a_4(\omega) e^{i\eta(\omega)}+e^{-i \mu (x-x')} a_3(\omega) e^{-i\eta(\omega)}\right]\;,\\
f_{3,\textrm{bulk}}^R(x,x',\omega) &=  -  \frac{ i \Delta_0 }{2v_F\sqrt{\Delta_0^2-\omega^2}}  e^{-\kappa(\omega)\lvert x -x' \rvert} \sin[\mu(x-x')]\;, \label{Eq:3_bulk} \\
f_{3,\textrm{interface}}^R(x,x',\omega) &=  -  \frac{\Delta_0 }{4v_F\sqrt{\Delta_0^2-\omega^2}}  e^{-\kappa(\omega)(x+x')}\left[  e^{i \mu (x-x')} a_4(\omega) e^{i\eta(\omega)}-e^{-i \mu (x-x')} a_3(\omega) e^{-i\eta(\omega)}\right]\;,\\
f_{+,\textrm{interface}}^R(x,x',\omega) &=  -  \frac{\Delta_0 }{2v_F\sqrt{\Delta_0^2-\omega^2}}  e^{-\kappa(\omega)(x+x')}  e^{i \mu (x+x')} b_3(\omega) \;,\\
f_{-,\textrm{interface}}^R(x,x',\omega) &=  -  \frac{\Delta_0 }{2v_F\sqrt{\Delta_0^2-\omega^2}}  e^{-\kappa(\omega)(x+x')} e^{-i \mu (x+x')} b_4(\omega) \;,
\end{align}
where $f_{+} = f_1 - i f_2$ (resp. $f_{-} = f_1 + i f_2$) is simply the $\ua\ua$ (resp. $\da\da$) pairing amplitude.  On the advanced side,
\begin{align}
f_{0,\textrm{bulk}}^A(x,x',\omega) &=  -  \frac{ \Delta_0 }{2v_F\sqrt{\Delta_0^2-\omega^2}}  e^{-\kappa(\omega)\lvert x -x' \rvert} \cos[\mu(x-x')]\;,\label{Eq:0_bulk_A}  \\
f_{0,\textrm{interface}}^A(x,x',\omega) &=  -  \frac{\Delta_0 }{4v_F\sqrt{\Delta_0^2-\omega^2}}  e^{-\kappa(\omega)(x+x')}\left[  e^{i \mu (x-x')} a_4^*(\omega) e^{i\eta(\omega)}+e^{-i \mu (x-x')} a_3^*(\omega) e^{-i\eta(\omega)}\right]\;,\\
f_{3,\textrm{bulk}}^A(x,x',\omega) &=  -  \frac{ i \Delta_0 }{2v_F\sqrt{\Delta_0^2-\omega^2}}  e^{-\kappa(\omega)\lvert x -x' \rvert} \sin[\mu(x-x')]\;,     \label{Eq:3_bulk_A}  \\
f_{3,\textrm{interface}}^A(x,x',\omega) &=  -  \frac{\Delta_0 }{4v_F\sqrt{\Delta_0^2-\omega^2}}  e^{-\kappa(\omega)(x+x')}\left[  e^{i \mu (x-x')} a_4^*(\omega) e^{i\eta(\omega)}-e^{-i \mu (x-x')} a_3^*(\omega) e^{-i\eta(\omega)}\right]\;,\\
f_{+,\textrm{interface}}^A(x,x',\omega) &=  -  \frac{\Delta_0 }{2v_F\sqrt{\Delta_0^2-\omega^2}}  e^{-\kappa(\omega)(x+x')} e^{i \mu (x+x')} b_4^*(\omega) \;,\\
f_{-,\textrm{interface}}^A(x,x',\omega) &=  -  \frac{\Delta_0 }{2v_F\sqrt{\Delta_0^2-\omega^2}}  e^{-\kappa(\omega)(x+x')} e^{-i \mu (x+x')} b_3^*(\omega)\;,
\end{align}
One can check on these expressions some of the affirmations made in the main text. In particular, the bulk contribution in $f_0$ is of the ESE type ($f_{0,\textrm{bulk}}^R(x,x',\omega) = f_{0,\textrm{bulk}}^R(x',x,\omega)$, while $f_{0,\textrm{bulk}}^R(x,x',\omega) = f_{0,\textrm{bulk}}^A(x',x,-\omega)$), and the bulk contribution is of the ETO type ($f_{3,\textrm{bulk}}^R(x,x',\omega) = - f_{3,\textrm{bulk}}^R(x',x,\omega)$, while $f_{3,\textrm{bulk}}^R(x,x',\omega) = f_{3,\textrm{bulk}}^A(x',x,-\omega)$). Furthermore, $f_{0,\textrm{interface}}$ gives rise to both an ESE and an OSO contribution, while $f_{3,\textrm{interface}}$ gives rise to both an OTE and an ETO contribution.

\subsection{D. Case of a ferromagnetic impurity}

So far, all expressions are completely general and we have not specified the elements of the scattering matrix. However, simple analytic expressions can be obtained in the case of a ferromagnetic impurity. We let $\delta \to 0$ and keep $m_1\delta = m_0$ finite. We find, below the gap

\begin{align}
b_1(\omega) &= \frac{1}{2i}\frac{\sinh(4 m_0)(e^{ -2i\lvert x_1 \rvert \omega}e^{i \eta(\omega)}+e^{ 2i\lvert x_1 \rvert \omega}e^{- i \eta(\omega)}) e^{i\lambda_1} }{e^{ 2i\lvert x_1 \rvert \omega}e^{- i \eta(\omega)}\sinh^2(2 m_0)+e^{ -2i\lvert x_1 \rvert \omega}e^{ i \eta(\omega)}\cosh^2(2 m_0)}\;, \label{Eq:ree}
\\
a_1(\omega) &= \frac{1}{e^{ 2i\lvert x_1 \rvert \omega}e^{- i \eta(\omega)}\sinh^2(2 m_0)+e^{ -2i\lvert x_1 \rvert \omega}e^{ i \eta(\omega)}\cosh^2(2 m_0)}\;, \label{Eq:reh} \\
b_3(\omega) &= \frac{1}{2i}\frac{\sinh(4 m_0)(e^{i \eta(\omega)}-e^{- i \eta(\omega)}) e^{-i\lambda_1} }{e^{ 2i\lvert x_1 \rvert \omega}e^{- i \eta(\omega)}\sinh^2(2 m_0)+e^{ -2i\lvert x_1 \rvert \omega}e^{ i \eta(\omega)}\cosh^2(2 m_0)}\;, \label{Eq:reep}
\\
a_3(\omega) &=- \frac{e^{ 2i\lvert x_1 \rvert \omega}\sinh^2(2 m_0)+e^{ -2i\lvert x_1 \rvert \omega}\cosh^2(2 m_0)}{e^{ 2i\lvert x_1 \rvert \omega}e^{- i \eta(\omega)}\sinh^2(2 m_0)+e^{ -2i\lvert x_1 \rvert \omega}e^{ i \eta(\omega)}\cosh^2(2 m_0)}\;. \label{Eq:rehp}
\end{align}
The other elements are given by particle-hole symmetry: $a_4(\omega) = - a_3(-\omega)^*$, $b_4(\omega) = - b_3(-\omega)^*$, $a_2(\omega) = - a_1(-\omega)^*$, $b_2(\omega) = - b_1(-\omega)^*$. The normal reflection coefficient vanishes for energies given by
\beq
4 \lvert x_1 \rvert \omega - 2\eta(\omega) = \pi + 2n\pi, \quad n\in \mathbb{Z}\;.
\eeq
Those are the energies of the bound-states that form between the ferromagnetic impurity and the superconductor. They induce perfect local Andreev reflection.

\end{document}